\documentclass[preprint]{vgtc}

\usepackage{amsmath}
\usepackage{booktabs}
\usepackage{tabularx}
\usepackage{listings}
\usepackage{tikz}
\usetikzlibrary{arrows.meta,positioning,fit,backgrounds}

\preprinttext{Accepted at the 2nd Workshop on GenAI, Agents, and the Future of VIS (VISxGenAI), IEEE VIS 2026}
\onlineid{1557}          
\vgtccategory{Research}
\vgtcpapertype{theory/model}  

\title{Measuring the Collaboration, Not the Agent:\\
An Evaluation Framework for Human--AI High-Dimensional Visual Analytics}

\author{Xinxiao Li\thanks{e-mail: xinxiaoli@info.shonan-it.ac.jp}\\%
  \scriptsize Faculty of Informatics, Shonan Institute of Technology\\%
  \scriptsize Tsujidonishikaigan, Fujisawa, Japan%
  \and Eisei Nakahara\thanks{Corresponding author. e-mail:
  nakahara.eisei@nihon-u.ac.jp}\\%
  \scriptsize College of Humanities and Sciences, Nihon University\\%
  \scriptsize Chiyoda, Tokyo, Japan}

\abstract{%
Benchmarks for visualization agents score the agent alone; the human only
supplies reference answers. For high-dimensional data this misses what matters:
the knowledge of which variables count is held by on-site experts, not by the
model, so the useful question is how well an analyst and an agent do
\emph{together}. We study one specific form of collaboration---the agent
proposes a view, the analyst accepts, rejects, or modifies it, and those
judgments are kept for later sessions---and propose five measures for it:
complementarity, intervention efficiency, appropriate reliance, knowledge
retention, and agency. The first four are computed from interaction logs alone.
We test the measures, not people: in a testbed for parallel-coordinates design
on public manufacturing data (UCI SECOM), with a simulated analyst whose
behavior we control, all four are computable and respond to the controlled
differences. Whether they reveal anything about real analysts is the next
step; we release the testbed for it.%
}

\keywords{Human--AI collaboration, visualization agents, evaluation,
provenance, parallel coordinates, high-dimensional data.}

\vgtcinsertpkg

\begin{document}

\firstsection{Introduction}
\maketitle
\label{sec:intro}

\textbf{In brief.} When an analyst and an AI agent design a visualization
together, how do we tell whether the pairing helped? We propose four log-based measures and
one survey-based measure of the pair, and show on a simulated analyst that the
four log-based ones can be computed and that they move when the collaboration
changes. We do \emph{not} show what they reveal about real people.

\textbf{Benchmarks measure the agent.}
Large multimodal models now produce valid and often reasonable charts from
natural-language requests, and an evaluation infrastructure has grown around
them quickly---VisEval for
NL-to-visualization validity~\cite{viseval2024}, nvBench~2.0 for ambiguity
resolution~\cite{nvbench2}, VisJudge-Bench for visualization quality
judging~\cite{visjudge2025}, DV-World for realistic multi-step
scenarios~\cite{dvworld2026}, and SciVisAgentBench for scientific
visualization agents~\cite{scivisagentbench2026}. All of these, however,
measure the AI in isolation: the human supplies reference answers or labels, or
is replaced by a user simulator~\cite{dvworld2026}. Collaborative systems
exist~\cite{hilsva2026, bonsai2026}, but \emph{building} one and \emph{showing
that the collaboration is what helps} are different problems, and for the
latter visualization has no yardstick.

\textbf{Why high-dimensional data needs one.}
For a scientific volume, a good view is largely fixed by the data's own
geometry, and the knowledge needed to find it is general enough to be in a
pretrained model. For high-dimensional abstract data---hundreds of anonymous
sensor channels, say---neither holds. Choosing which few of hundreds of
variables to show, in what order, is a combinatorial choice with no geometric
anchor; whether a view is good depends on the question asked; and the knowledge
of which sensors matter for \emph{this} plant lives with its
engineers~\cite{sdpcPrior}, not in any model. The agent can search the space
and the expert can judge it; each holds part of what is needed, so performance
is a property of the pair. And because that expert knowledge is missing from the
model, the agent should \emph{keep} what it is told rather than be retaught
every session, which raises a question agent-only benchmarks never ask: does
the agent need less correction over time? Collaboration also cannot be assumed
to help---AI suggestions can fixate designers and narrow
exploration~\cite{fixation2024, guidance2025}---so it has to be measured.

\textbf{Which collaboration.}
We study one deliberately narrow form: the agent \emph{proposes} a view, the
analyst \emph{accepts, rejects, or modifies} it, and the analyst's judgments are
\emph{kept} and fed back to the agent in later sessions. This is persisted
corrective feedback, not open-ended co-creation or joint brainstorming; we use
``collaboration'' in this sense throughout.

\textbf{Contributions.}
(1)~An argument for measuring the analyst--agent pair in high-dimensional
data (\S\ref{sec:problemclass}). (2)~Five measures (``axes'') of the pair; four
are computed from interaction logs (\S\ref{sec:framework}). (3)~A testbed
showing those four are computable and respond to controlled variation
(\S\ref{sec:testbed}--\ref{sec:results})---validation as \emph{instruments},
with a simulated analyst, not a claim about people. (4)~The testbed and log
format, released for human studies.

\section{Background}
\label{sec:background}

\textbf{Visualization agents and collaborative systems.}
Beyond the benchmarks above, evaluation-centric work taxonomizes scientific
visualization agents~\cite{evalcentric2025}; none measures the human--agent
\emph{team}. Collaborative systems such as HiLSVA~\cite{hilsva2026} and
BONSAI~\cite{bonsai2026} build on paired recommendation, proactive
intervention, and multimodal intent~\cite{haichart2024, proactiveva2025,
interchat2025}, descending from semantic interaction~\cite{endert2012}. Recent
work organizes agentic visual
analytics into roles and workflows~\cite{coevolution2026} and finds persistent
agent memory helps only insofar as retained feedback is of high
quality~\cite{interactionparadigms2026}---foreshadowing our retention axis.

\textbf{Collaboration measurement elsewhere.}
Human--AI teaming supplies constructs visualization has not imported:
\emph{complementarity}, requiring team loss strictly below both solo
losses~\cite{complementarity2024}; \emph{appropriate reliance}, separating
justified from blind acceptance~\cite{reliance2026, decisiontheoretic2024};
\emph{readiness}, whose learning-over-time family is to our knowledge the only
existing construct for longitudinal knowledge transfer in a
team~\cite{readiness2026}; and frameworks for selecting metrics by collaboration
mode~\cite{fragiadakis2024}. Outside visualization, team-level benchmarks
have begun to appear: CentaurEval builds coding tasks that neither humans nor
LLMs solve alone and compares the team against both~\cite{centaureval2026},
and Collaborative Gym evaluates agents that work with real or simulated users
on both outcome and process~\cite{cogym2024}. Neither addresses visualization
design, and, to our knowledge, neither measures whether the agent retains what the human taught
it across sessions---the axis our problem class makes central. None of these
constructs has an operational counterpart in visualization-agent evaluation.

\section{Why the Problem Class Matters}
\label{sec:problemclass}

Why insist on collaboration-level evaluation \emph{here}---in high-dimensional
abstract data---rather than wherever agents are strongest? Because the need
follows from the structure of the problem class.
Table~\ref{tab:problemclass} contrasts scientific visualization, where agent
benchmarks are most developed, with high-dimensional abstract data, exemplified
by parallel-coordinates (PCP) design over multivariate process data.

\begin{table}[tb]
  \caption{Structural differences between the problem classes. Each row
  feeds the next; the final row is the premise of our framework.}
  \label{tab:problemclass}
  \scriptsize
  \centering
  \begin{tabular}{@{}p{0.21\linewidth}p{0.33\linewidth}p{0.34\linewidth}@{}}
    \toprule
    & \textbf{Scientific vis.} & \textbf{High-dim.\ abstract data} \\
    \midrule
    Design space & Continuous (camera, transfer function, isovalue) &
    \textbf{Combinatorial} (axis subset, ordering, inversion, scaling) \\
    Grounding of ``good'' & Spatial referent in the data itself &
    \textbf{Task-relative only} \\
    Expertise & Largely domain-general; plausibly present in pretrained
    models & \textbf{Site-specific, non-transferable} \\
    Interaction semantics & Navigation & \textbf{Direct manipulation =
    design decision} \\
    $\Rightarrow$ Feedback lifecycle & Test-time adaptation suffices &
    \textbf{Must be distilled and retained} \\
    \bottomrule
  \end{tabular}
\end{table}

The rows are causally ordered, and each one dictates something the evaluation
must measure. A combinatorial space cannot be searched by the expert alone,
yet without a spatial referent the agent cannot tell a good view from a bad
one; so the question is whether the \emph{pair} beats each member
(axis~1). Expert attention is the scarce resource in that search, so its cost
must be counted (axis~2). Since the expertise is site-specific---which
variables \emph{matter in this plant, this line, this week}~\cite{sdpcPrior}---the
agent's proposals cannot be checked against general knowledge, and whether the
analyst accepts them for good reasons becomes a question in itself (axis~3).
Because that expertise is absent from the model, it must be kept across
sessions, so whether the agent needs less correction over time is measurable
and matters (axis~4). Finally, it is expressed by direct
manipulation---reordering an axis \emph{is} a design
judgment~\cite{endert2012}---so the interaction log is where all of this
becomes observable.

\section{The Five-Axis Framework}
\label{sec:framework}

Table~\ref{tab:framework} summarizes the framework: each axis answers one
question about the team under an operational definition over concrete data.

\begin{table*}[tb]
  \caption{The five-axis framework. Axes~1--4 are computed from interaction
  logs; axis~5 requires post-session questionnaires.}
  \label{tab:framework}
  \scriptsize
  \centering
  \begin{tabularx}{\textwidth}{@{}%
    >{\raggedright\arraybackslash}p{0.21\textwidth}%
    >{\raggedright\arraybackslash}p{0.17\textwidth}%
    >{\raggedright\arraybackslash}X%
    >{\raggedright\arraybackslash}p{0.14\textwidth}@{}}
    \toprule
    \textbf{Axis} & \textbf{Question} & \textbf{Operationalization} &
    \textbf{Primary data} \\
    \midrule
    1. Attainment \& complementarity & Does the team beat both
    solos? & $L(\mathit{team}) < \min\bigl(L(\mathit{human}),
    L(\mathit{AI})\bigr)$ & interaction log \\
    2. Efficiency \& intervention budget & At what human cost? &
    interventions-to-attainment; marginal gain per intervention
    $\Delta L / \Delta I$ & interaction log \\
    3. Appropriate reliance \& fixation & Blind following?
    Narrowed search? & acceptance rate split by proposal correctness;
    design-space coverage & interaction log \\
    4. Knowledge retention & Does the agent need less correction over time? &
    $\Delta$ expert-alignment after distillation; interventions needed in
    session $n{+}1$ & logs across sessions \\
    5. Agency, trust, load & Will experts keep using it? &
    CSI, trust calibration, NASA-TLX & questionnaire \\
    \bottomrule
  \end{tabularx}
\end{table*}

\textbf{Axes 1--3.}
Table~\ref{tab:framework} carries the operational definitions; three points
need stating. With $L(c)$ the task loss under condition $c$, complementarity
requires $L(\mathit{team}) < \min(L(\mathit{human}),
L(\mathit{AI}))$~\cite{complementarity2024}---deliberately strict, since a team
merely matching the better solo member does not justify its coordination cost.
Efficiency counts intervention events (rejections, manual reorderings, axis
substitutions) and reports interventions-to-attainment alongside marginal
improvement per intervention $\Delta L / \Delta I$, a budget--performance curve
whose saturation characterizes how far human attention goes; attainment alone
hides human cost. Reliance splits the analyst's acceptance rate by whether the
proposal was in fact correct, separating over-reliance (accepting wrong
proposals) from under-reliance (rejecting right ones)~\cite{reliance2026,
decisiontheoretic2024}, paired with design-space coverage---axis-subset
diversity relative to trials spent---to capture fixation~\cite{fixation2024}.

\textbf{Axis 4: Knowledge retention.}
The distinguishing axis for our problem class (\S\ref{sec:problemclass}) and the
visualization counterpart of readiness' learning-over-time
family~\cite{readiness2026}: after distilling one session's accepted and
rejected judgments into the agent, does the \emph{next} session need fewer
interventions to reach the same attainment? We measure session $n{+}1$ with and
without distillation.

\textbf{Axis 5: Agency, trust, and load.}
Sustained adoption depends on the expert's sense of agency and calibrated
trust; we adopt standard instruments (CSI, trust calibration, NASA-TLX). This
axis requires human participants and is out of scope for the pilot.

\section{Testbed and Pilot Study}
\label{sec:testbed}

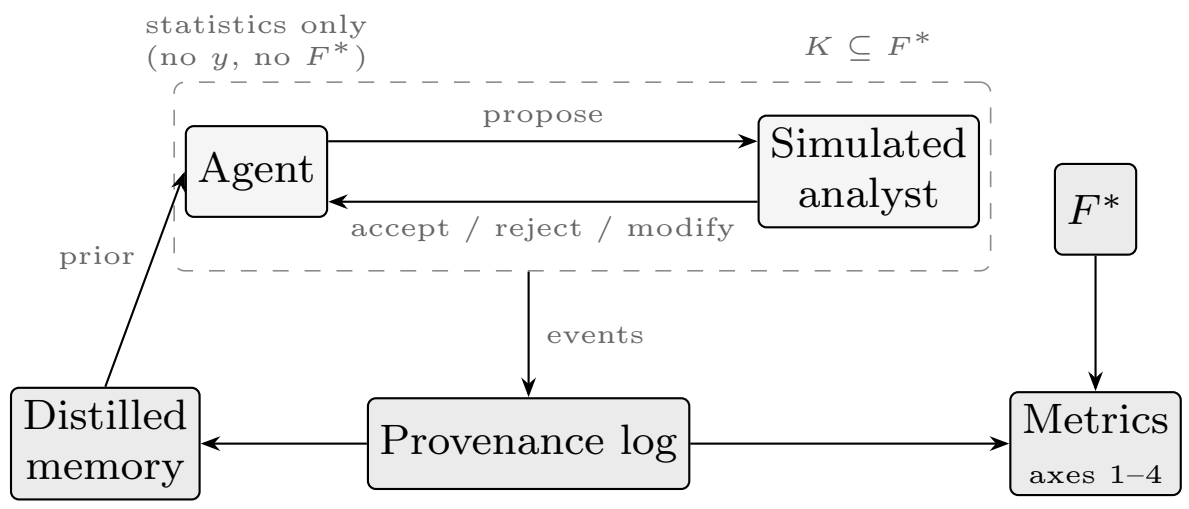
\begin{figure}[tb]
  \centering
  \begin{tikzpicture}[
    font=\scriptsize,
    box/.style={draw, rounded corners=1.5pt, align=center, inner sep=2.2pt,
                minimum height=6mm, line width=0.4pt},
    actor/.style={box, fill=black!4},
    store/.style={box, fill=black!8},
    flow/.style={-{Stealth[length=1.3mm]}, line width=0.4pt},
    note/.style={font=\tiny, text=black!60},
  ]
    \node[actor] (agent)   at (0.2,0)      {Agent};
    \node[actor] (analyst) at (4.25,0)     {Simulated\\analyst};
    \node[store] (mem)     at (-0.8,-1.8)  {Distilled\\memory};
    \node[store] (log)     at (2.0,-1.8)   {Provenance log};
    \node[store] (metrics) at (5.75,-1.8)  {Metrics\\\tiny axes~1--4};
    \node[store] (fstar)   at (5.75,-0.25) {$F^{*}$};

    \draw[flow] ([yshift=2.0mm]agent.east) --
      node[above=-0.2mm, note] (prop) {propose}
      ([yshift=2.0mm]analyst.west);
    \draw[flow] ([yshift=-2.0mm]analyst.west) --
      node[below=-0.2mm, note] (fb) {accept / reject / modify}
      ([yshift=-2.0mm]agent.east);

    \node[note, above=2.4mm of agent, align=center]
      {statistics only\\(no $y$, no $F^{*}$)};
    \node[note, above=2.4mm of analyst, align=center]
      {$K \subseteq F^{*}$};

    \begin{scope}[on background layer]
      \node[draw, dashed, line width=0.3pt, rounded corners=2pt, black!45,
            fit=(agent)(analyst)(prop)(fb), inner sep=2pt] (loop) {};
    \end{scope}

    \draw[flow] (loop.south -| log.north) --
      node[right, note] {events} (log.north);
    \draw[flow] (log.east) -- (metrics.west);
    \draw[flow] (fstar.south) -- (metrics.north);
    \draw[flow] (log.west) -- (mem.east);
    \draw[flow] (mem.north) -- node[left, note, pos=0.6] {prior} (agent.west);
  \end{tikzpicture}
  \caption{Testbed architecture. The dashed region is one session. Neither actor
  sees what the other knows---the agent has summary statistics and correlation
  structure but no labels and no $F^{*}$, the analyst only $K \subseteq F^{*}$
  and a noisy percept of render quality. All four axes are computed downstream
  from the event stream (\S\ref{sec:schema}) and $F^{*}$ alone, never from
  simulator state, so the metric code runs unchanged on a human-study log.}
  \label{fig:architecture}
\end{figure}

The testbed implements the propose--respond loop for PCP design
(Fig.~\ref{fig:architecture}): an agent proposes candidate views, a simulated
analyst accepts, rejects, or modifies them, and both actors' actions are
logged. Axes~1--4 are then computed from that log. Its purpose is to check the
\emph{measures}: because we set the simulated analyst's behavior, we know what
each measure ought to report and can check that it does.

\subsection{Task and data}
\label{sec:task}

We use the UCI SECOM dataset~\cite{secom}: 1{,}567 production runs from a
semiconductor line, 590 sensor features, pass/fail labels. The task is to
identify, via PCP inspection, the feature subset discriminating failures. Since
true physical causes are unknown, we \emph{operationally define} ground truth
as a subset $F^{*}$ derived beforehand by supervised analysis: stability
selection with $\ell_1$-regularized logistic regression over 100 bootstrap
resamples, then greedy forward selection on nested cross-validated AUC, giving
$|F^{*}| = 8$ features at AUC $0.705 \pm 0.047$---a proxy, not physical truth
(\S\ref{sec:agenda}).
Preprocessing removes columns more than 50\% missing, constant and
quasi-constant columns, and one side of every feature pair with $|r| > 0.95$,
leaving $n = 272$ of the 590 features over 1{,}567 runs with 104 failures
(6.6\%).

\subsection{Design space and agent}
\label{sec:designspace}

A candidate PCP view is a tuple $c = (S, \pi, \mathbf{f}, \mathbf{s})$: an axis
subset $S$ with $|S| = k \in [6, 10]$, an ordering $\pi$, per-axis inversion
flags $\mathbf{f} \in \{0,1\}^{k}$, and per-axis scalings $\mathbf{s} \in
\{\text{linear}, \text{log}, \text{rank}\}^{k}$, so the design space has size
$\binom{n}{k} \, k! \, 2^{k} 3^{k}$---approximately $4.5 \times 10^{25}$
candidates at $n = 272$, $k = 8$---the concrete instance of
Table~\ref{tab:problemclass}'s first row.
The agent is deliberately two-layered. Every headline number below comes from a
rule-based proposer that samples axes from label-free structure heuristics
(variance, outlier rate, correlation centrality, Laplacian score) then
hill-climbs order, orientation, and scale on a label-free readability score;
running $\sim$5{,}000 sessions through a language-model API is not tractable. A
prompt-based LLM proposer (\texttt{gpt-5.4-mini}) implements the identical
interface, and \S\ref{sec:results} reports a three-cell replication. Because
SECOM sensor names are anonymous numeric identifiers, a pretrained model has no
semantic prior to exploit---itself an instance of the argument in
\S\ref{sec:problemclass} that site-specific knowledge of \emph{which} variable
matters is exactly what pretraining does not supply.

\subsection{Simulated analyst}
\label{sec:simulant}

The simulated analyst is not a substitute for participants. Unlike a simulator
standing in for the user being evaluated~\cite{dvworld2026}, it does not
replace the object of measurement; it calibrates the instrument.

The policy has three parameters in $[0,1]$: \texttt{expertise} (fraction of
$F^{*}$ known), \texttt{reliance} $\rho$ (propensity to accept proposals), and
\texttt{exploration} (propensity to try candidates beyond proposals). Sweeping
\texttt{reliance} is what lets axis~3 separate over- from
under-reliance---the pilot's central device.

Five conditions; no human takes part in any of them. \textbf{C0}, a sanity
check pairing random proposals with an accept/reject-only analyst---if this
``team'' scored as complementary, the metric would be broken; \textbf{C1}
analyst-solo (the simulated analyst with no agent proposals), \textbf{C2}
agent-solo (agent plus automated judge, no analyst), \textbf{C3}
collaboration (propose--accept/reject/modify), and \textbf{C3$'$} collaboration
whose judgments are \emph{not} carried to the next session (ablation for
axis~4).

\subsection{Provenance schema}
\label{sec:schema}

Every condition emits one event stream:

\begin{lstlisting}
event = {t: float, session_id: str, trial: int,
  actor: "agent" | "analyst" | "judge",
  action: "propose" | "accept" | "reject" | "reorder"
        | "drop" | "add" | "scale" | "flip" | "brush"
        | "filter" | "conclude",
  target: {...}, params: {...},
  candidate_id: str, meta: {...}}
\end{lstlisting}

Put plainly, axes~1--4 are computed from user logs: the metric code reads only
these events (plus $F^{*}$ to score loss), never the simulator's internal
state, so it runs unchanged on logs from a human study. The \texttt{meta}
field, which holds simulator internals such as perceived quality, is excluded
for the same reason. The same accept/reject events are also what the agent
distills into memory for the next session, so logging serves both evaluation
and learning.

\begin{figure}[tb]
  \centering
  \includegraphics[width=\linewidth]{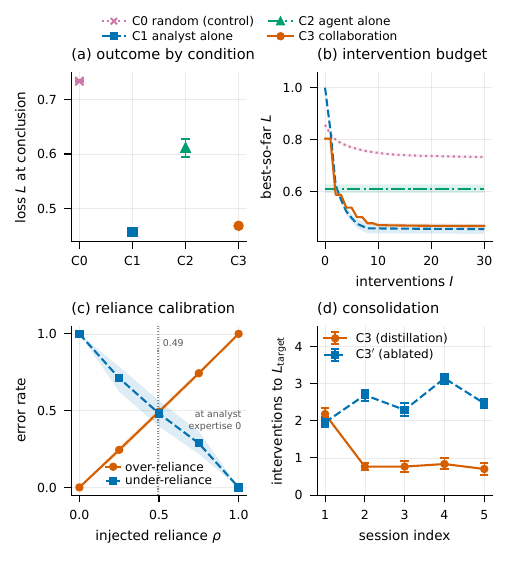}
  \caption{Pilot results for axes~1--4 (values in Table~\ref{tab:results}).
  (a) $L$ at conclusion by condition; marginals over different parameter
  cells, so not a ranking. (b) running-best $L$ against interventions spent;
  C2 is a reference level. (c) over- and under-reliance (both errors) against
  the injected acceptance propensity $\rho$ at \texttt{expertise}${}=0$.
  (d) interventions to reach $L_{\mathrm{target}}$ across five sessions, with
  (C3) and without (C3$'$) carried-over judgments. Intervals are seed-to-seed
  variation in the simulator, not inference about people.}
  \label{fig:results}
\end{figure}

\begin{table*}[tb]
  \caption{Pilot results (simulated analyst). $L$ is a loss: lower is better.
  Brackets: bootstrap 95\% CIs over seeds. ``---'': undefined for that
  condition (C1 has no proposals to rely on; C2 spends no interventions). The
  last column is the evidence that the axis discriminates.}
  \label{tab:results}
  \scriptsize
  \centering
  \setlength{\tabcolsep}{3.5pt}
  \begin{tabularx}{\textwidth}{@{}%
    >{\raggedright\arraybackslash}p{0.235\textwidth}%
    llll%
    >{\raggedright\arraybackslash}X@{}}
    \toprule
    \textbf{Axis: measure (from the log)} & \textbf{C0 random} &
    \textbf{C1 analyst-solo} & \textbf{C2 agent-solo} & \textbf{C3 collab.} &
    \textbf{Evidence of discrimination} \\
    \midrule
    1: mean $L$ at conclusion & .734 [.732,.736] & .456 [.454,.459] &
    .612 [.595,.628] & .468 [.466,.470] & all four CIs disjoint \\
    1: grid cells where team beats both solos & 0\% & --- & --- & 34.7\% &
    C3 by \texttt{expertise} 0/.25/.5/.75/1: 93/80/0/0/0\% \\
    2: $\Delta L/\Delta I$ over first 5 interventions & .013 & .085 & --- &
    .066 & return per intervention differs by condition \\
    2: interventions until the curve saturates & 16 & 6 & --- & 8 &
    sharp knee; C0 saturates late at a poor loss \\
    3: over-/under-reliance as $\rho$ goes $0{\to}1$ & --- & --- & --- &
    $0{\to}1$ / $1{\to}0$ & cross at $\rho = 0.49$: injected reliance read back \\
    3: distinct axis sets / share of features seen & 29.9 / 59\% & 7.4 / 2.6\%
    & 5.7 / 13\% & 19.1 / 18\% & analyst-solo search is narrow (fixation) \\
    4: interventions to $L_{\mathrm{target}}$, session 1$\to$5 & --- & --- &
    --- & 2.18$\to$0.70 & C3$'$: 1.95$\to$2.47; slope diff.\ $-0.44$
    [$-0.50,-0.37$] \\
    \bottomrule
  \end{tabularx}
\end{table*}

\section{Results}
\label{sec:results}


For each axis we ask two questions: can it be computed from the log, and does
it separate conditions that we know, by construction, to differ? We make no
claim that collaboration helps people; a simulator cannot establish that.
The grid crosses \texttt{expertise} and \texttt{reliance} $\in \{0, .25, .5,
.75, 1\}$ with \texttt{exploration} $\in \{.1, .3, .5\}$ over 30 seeds, 30
trials per session and a hard budget of 30 interventions: 4{,}980 sessions,
120{,}755 distinct designs, 177\,s on one core. The attainment threshold
$L_{\mathrm{target}} = 0.624$ was fixed in advance from a separate pilot sweep
(seeds 0--4, excluded here). We report bootstrap 95\% CIs over seeds and no
$p$-values, since on a simulator sample size is a free parameter.
Table~\ref{tab:results} collects the numbers; the text gives the takeaway per
axis.

\textbf{Axis 1: complementarity is computable and has a visible boundary.}
The measure is the loss of the best design the analyst endorsed. All four
condition means separate cleanly (Table~\ref{tab:results};
Fig.~\ref{fig:results}a), but they are not a ranking: they average over
different parameter cells, and a fifth of C3's cells sit at
\texttt{reliance}${}=1$, where the analyst accepts the first proposal and is
effectively C2 (dropping them moves C3 to $0.436$, below C1). We therefore
compare cell by cell. The team beats both solo baselines in $34.7\%$ of cells,
all at low analyst expertise; at \texttt{expertise}${}\geq 0.5$ the simulated
analyst does better alone---a boundary we report rather than suppress. Part
of it is built in (an analyst holding all of $F^{*}$ knows exactly what the
loss scores), and part is a mechanism worth testing with people: the agent's
tidy layouts raise perceived quality, so the analyst stops early. The
random-proposal control C0 is never complementary, so the effect is not a
metric artifact.

\textbf{Axis 2: the cost of human attention has a measurable knee.}
The measure is the running-best loss against interventions spent. C1 and C3
gain most from the first few and saturate sharply (at 6 and 8); C0 gains
little and saturates late (Fig.~\ref{fig:results}b). The curve has a readable
knee, so ``how much analyst effort did this team need'' has a concrete answer.

\textbf{Axis 3: the measure reads back the reliance we injected.}
The measure splits acceptances by whether the proposal was in fact good
(below $L_{\mathrm{target}}$; $52\%$ were). Sweeping the analyst's acceptance
propensity $\rho$, over-reliance (accepting bad proposals) rises from 0 to 1
while under-reliance (rejecting good ones) falls from 1 to 0
(Fig.~\ref{fig:results}c). That the curves follow $\rho$ is the intended
calibration, not circularity: the instrument recovers a disposition whose true
value we set. The separation is cleanest without analyst expertise; with
expertise, under-reliance barely varies with $\rho$ (its range shrinks to
$0.09$--$0.14$), so the separation is itself expertise-dependent. Coverage adds a fixation signal visible in the log alone:
analyst-solo search revisits a small familiar set of axes.

\textbf{Axis 4: retention separates from mere drift only with the ablation.}
The measure is the number of interventions needed to reach
$L_{\mathrm{target}}$ in each of five successive sessions, run on successive
time blocks of the data. With the analyst's judgments carried forward (C3)
this falls from $2.18$ to $0.70$; without them (C3$'$) it does not fall
(Fig.~\ref{fig:results}d), and the difference in slopes has a CI well below
zero. The ablation is necessary: blocks differ in failure count (44, 21,
11, 11, 17), so a falling C3 curve alone could mean easier blocks.

\textbf{Robustness.}
With an alternative ground truth chosen by mutual information, sharing no
features with $F^{*}$, the axes still separate the conditions. Reweighting the
quality score keeps that separation but can flip C1 versus C3: weighting
readability and scale at $0.60$ instead of $0.40$ puts C3 ahead and doubles the
complementary region to $67\%$ of cells, so whether collaboration ``wins''
partly depends on the weights. The agent's label-free heuristics recover 2 of
the 8 $F^{*}$ features (chance: $0.23$), so C2 is a real baseline.

\textbf{The measures do not depend on the agent.}
Replacing the rule-based proposer with an LLM (\texttt{gpt-5.4-mini}) on three
cells $\times$ five seeds leaves every axis computable and orders the cells
identically (rank agreement $1.0$, no malformed proposals; mean loss $0.469$
vs.\ $0.421$). Responses are cached by prompt (49 distinct prompts in 15
sessions), so seed variance there is understated; this is not a comparison of
agents.

\section{Limitations and Research Agenda}
\label{sec:agenda}

\textbf{What this paper shows, and what it does not.}
Four measures of analyst--agent collaboration can be computed from logs and
respond correctly to differences we control; whether they say anything useful
about real analysts, who anchor, fixate, and learn, is untested.
(1)~No human participants: axis~5 is unvalidated and axes~1--4 are validated
only as \emph{instruments}, not as findings about people. (2)~SECOM's $F^{*}$ is
a derived proxy, so attainment means agreement with a supervised analysis.
(3)~The simulated analyst's parameters are design choices, not fitted to human
data. (4)~Collaboration here means persisted corrective feedback; richer forms
such as joint exploration or dialogue would need further measures.

\textbf{Agenda.}
The next step is a human study using the testbed's pre-specified metrics.
\emph{Which axis should be validated with experts first?} Complementarity is
the headline construct; retention is the one agent-only benchmarks cannot see.



\bibliographystyle{abbrv-doi}
\bibliography{references}

\end{document}